\def\year{2022}\relax
\documentclass[letterpaper]{article} 
\usepackage{aaai22}  
\usepackage{times}  
\usepackage{helvet}  
\usepackage{courier}  
\usepackage[hyphens]{url}  
\usepackage{graphicx} 
\usepackage{natbib}  
\usepackage{caption} 
\DeclareCaptionStyle{ruled}{labelfont=normalfont,labelsep=colon,strut=off} 
\usepackage{algorithm}
\usepackage{algorithmic}

\usepackage{newfloat}
\usepackage{listings}
\floatstyle{ruled}
\newfloat{listing}{tb}{lst}{}
\floatname{listing}{Listing}
\usepackage{multirow}
\usepackage{listings}
\usepackage{color,xcolor}

\definecolor{lightgray}{HTML}{F4F4F7}
\definecolor{darkgray}{rgb}{.4,.4,.4}
\definecolor{purple}{rgb}{0.65, 0.12, 0.82}
\lstdefinelanguage{JavaScript}{
  keywords={break, case, catch, continue, debugger, default, delete, do, else, false, finally, for, function, if, in, instanceof, new, null, return, switch, this, throw, true, try, typeof, var, void, while, with, async, await},
  morecomment=[l]{//},
  morecomment=[s]{/*}{*/},
  morestring=[b]',
  morestring=[b]",
  ndkeywords={class, export, boolean, throw, implements, import, this},
  keywordstyle=\color{blue}\bfseries,
  ndkeywordstyle=\color{darkgray}\bfseries,
  identifierstyle=\color{black},
  commentstyle=\color{purple}\ttfamily,
  stringstyle=\color{red}\ttfamily,
  sensitive=true,
  showspaces=false,               
  showstringspaces=false,         
  showtabs=false,                 
}

\usepackage{xcolor}
\usepackage{subfig}
\newif\ifcomment
\commentfalse
\ifcomment
\newcommand{\sayak}[1]{{\bf \textcolor{purple}{Sayak: #1}}}
\newcommand{\dipanjan}[1]{{\bf \textcolor{red}{Dipanjan: #1}}}
\newcommand{\priyanka}[1]{{\bf \textcolor{orange}{Priyanka: #1}}}
\newcommand{\giovanni}[1]{{\bf \textcolor{orange}{Giovanni: #1}}}
\newcommand{\chris}[1]{{\bf \textcolor{orange}{Chris: #1}}}
\newcommand{\shirin}[1]{{\bf \textcolor{brown}{Shirin: #1}}}

\newcommand{\change}[1]{{\bf \textcolor{blue}{#1}}}
\else
\newcommand{\shirin}[1]{}
\newcommand{\change}[1]{}
\newcommand{\sayak}[1]{}
\newcommand{\dipanjan}[1]{}
\newcommand{\priyanka}[1]{}
\newcommand{\giovanni}[1]{}
\newcommand{\chris}[1]{}
\fi

\newif\ifadditions
\additionstrue

\ifadditions
\newcommand{\additions}[1]{{\bf \textcolor{blue}{#1}}}
\else
\newcommand{\additions}[1]{}
\fi

\title{Demis}
\title{Unveiling the Risks of NFT Promotion Scams}

\author{
    Sayak Saha Roy\textsuperscript{\rm 1}, 
    Dipanjan Das\textsuperscript{\rm 2}, 
    Priyanka Bose\textsuperscript{\rm 2}, 
    Christopher Kruegel\textsuperscript{\rm 2}, 
    Giovanni Vigna\textsuperscript{\rm 2}, 
    Shirin Nilizadeh\textsuperscript{\rm 1}
}
\affiliations{
    \textsuperscript{\rm 1}The University of Texas at Arlington, Texas, USA\\
    sayak.saharoy@mavs.uta.edu, shirin.nilizadeh@uta.edu \\
    \textsuperscript{\rm 2}University of California, Santa Barbara, California, USA \\
    \{dipanjan, priyanka, chris, vigna\}@cs.ucsb.edu
}

\begin{document}
\maketitle
\begin{abstract}
\sayak{Change: Rewritten to focus on findings}
The rapid growth in popularity and hype surrounding digital assets such as art, video, and music in the form of non-fungible tokens (NFTs) has made them a lucrative investment opportunity, with NFT-based sales surpassing \$25B in 2021 alone. 
However, the volatility and general lack of technical understanding of the NFT ecosystem have led to the spread of various scams.
The success of an NFT heavily depends on its online virality.
As a result, creators use dedicated promotion services to drive engagement to their projects on social media websites, such as Twitter. 
However, these services are also utilized by scammers to promote fraudulent projects that attempt to steal users' cryptocurrency assets, thus posing a major threat to the ecosystem of NFT sales. 

In this paper, we conduct a longitudinal study of 439 promotion services (accounts) on Twitter that have collectively promoted 823 unique NFT projects through giveaway competitions over a period of two months.
Our findings reveal that more than 36\% of these projects were fraudulent, comprising of phishing, rug pull, and pre-mint scams. We also found that a majority of accounts engaging with these promotions (including those for fraudulent NFT projects) are bots that artificially inflate the popularity of the fraudulent NFT collections by increasing their likes, followers, and retweet counts. 
This manipulation results in significant engagement from real users, who then invest in these scams.
We also identify several shortcomings in existing anti-scam measures, such as blocklists, browser protection tools, and domain hosting services, in detecting NFT-based scams.
We utilize our findings to develop a machine learning classifier tool that was able to proactively detect 382 new fraudulent NFT projects on Twitter.


\end{abstract}

\section{Introduction}

Non-fungible tokens (NFTs) are digital assets such as art, videos, music, etc., which can be tracked using a unique identifier stored on a blockchain.
Blockchains are decentralized ledgers that keep a comprehensive record of how these assets are traded. 
NFTs can only be purchased or exchanged using a specific cryptocurrency, such as Ethereum, Wrapped Ether (WETH), etc. 
The decentralized nature of this system provides the convenience and transparency of owning and transferring digital assets. 
This combined with an assumption of increased protection against theft and unlawful distribution compared to other digital goods on the Internet has led to NFTs gaining widespread popularity in recent years, hitting sales of 25 billion dollars in 2021 alone~\cite{reuters_nft_2021}. 
Similar to other digital goods, NFTs can be purchased on dedicated marketplaces such as OpenSea~\cite{opensea}, though they are also sold independently by the creator on their website.
Factors that heavily impact the popularity of a specific NFT project include endorsements from celebrity figures~\cite{nadini2021mapping}, digital representation of popular culture~\cite{thompson2021untold}, utility -- such as giving access to exclusive events or games~\cite{decentraland}. 

Not surprisingly, the hype around NFTs has attracted scammers~\cite{atzori_ozsoy_2022} who seek to take advantage of unsuspecting investors.
Some of the common scams in this ecosystem include NFTs that claim to be created by famous artists when they are actually a copy, forgery, or phishing scam~\cite{kshetri2022scams}. 
In other cases, NFT projects also claim to be ``rare'' or part of a ``limited edition,'' promising huge monetary returns in the future, with the creator abandoning the project entirely and disappearing with the profit from initial sales~\cite{nft_rug_pull_evolved_apes}, a strategy also termed as \textit{rug pull}~\cite{das2021understanding}). 
To ensure these scams reach a large audience, there have been recent reports of attackers hiring social media influencers or online ``shills'' to promote their NFTs~\cite{decrypt_rick_2021}. 
These tactics can create an illusion of value, demand, and \textit{hype} for the fraudulent NFT, leading to financial losses for those who invest in these assets. Some reports claim that more than \$100M was lost due to NFT scams in 2021~\cite{decrypt_cybercriminals_2022}. 

While prior research has studied the features of NFT forgery and rug pull scams~\cite{das2021understanding}, to the best of our knowledge, there  exists no effort that measures the characteristics and negative impact of the \emph{promotion} of fraudulent NFT projects on social media. In light of this, we aim to answer the following three research questions:
\sayak{Change: Only 3 RQs now instead of 6}
\textbf{RQ1: }What are the characteristics and visibility of fraudulent NFT accounts that run promotions on Twitter?
\textbf{RQ2: }How effective are prevalent anti-scam measures towards detecting these attacks?
\textbf{RQ3: }What is the financial impact of these scams?

Our work closely monitors 439 Twitter accounts that promoted 823 unique NFT projects through incentive-driven \textit{giveaway} competitions from June 15th to August 20th, 2022, making the following contributions to  provide a comprehensive understanding of these scams:
\textbf{i)}~\textbf{Characterizing} how such promotions utilize artificial followers and engagement to ``hype'' the popularity of NFT projects, including those which are fraudulent. 
\textbf{ii)}~\textbf{Investigating} the efficiency of spam blocklists and browser protection tools against NFT-based attacks, identifying crucial gaps in the prevalent NFT ecosystem in the process.
\textbf{iii)}~\textbf{Tracking} the transactions of the fraudulent NFT projects to identify the financial damage done by the NFT scams that are promoted 
\textbf{iv)}~\textbf{Building a machine-learning-based model} that was able to proactively find 382 new fraudulent projects on Twitter.

\section{Background and Related Work}

\subsubsection{The Blockchains ecosystem:} 
Blockchains~\cite{ibm_blockchain} are digital ledgers that are decentralized in nature, with a large network of computing systems consistently validating and adding new records, making them highly resistant to malicious modification and fraud.
Blockchains served as the foundation for cryptocurrencies, establishing a decentralized and secure environment for digital transactions. More recently, these innovative networks also power Non-Fungible tokens (NFTs), digital goods that can be traded using cryptocurrency assets.
Popular blockchains used for creating and trading NFTs include Ethereum~\cite{ethereum} and the Binance Smart Chain~\cite{binance_chain}. Although these blockchains aim to provide transparency and security, their adoption has also paved the way for various illicit activities. Prior literature has extensively studied scams that target cryptocurrency assets such as fake mining/giveaway scams that impersonate popular crypto websites and false donation campaigns~\cite{lidouble,wu2020phishers}, ransomware attacks~\cite{alqahtani2022survey} and Ponzi schemes~\cite{vasek2019analyzing,}. Deceptive crypto services such as fake exchanges~\cite{amiram2020competition}, scam wallets~\cite{vasek2015there,dumitrescu2017bitcoin} are also prevalent. These scams are often difficult to regulate and significantly impact the crypto sphere due to the pseudonymous and irreversible nature of blockchain transactions~\cite{wu2021analysis}. However, prior literature on NFT-based scams has been sparse, with most work focused on characterizing isolated fraudulent projects from NFT Marketplaces~\cite{das2021understanding,kshetri2022scams}. In this work, we identify the impact of social media - one of the primary factors which decide the virality of NFT projects, towards making NFT-based scams more visible, and also develop countermeasures to detect them.

\subsubsection{Overview on NFTs:} NFTs, or non-fungible tokens, have gained significant popularity in the digital art and collectibles space. NFT creators often group multiple assets under a project or "collection" based on a specific theme. Each collection is assigned a unique "contract address" that enables transactions, such as buying and selling, for all NFTs within that collection. The blockchain records transaction history, ownership, and other relevant information.
To manage the assets within an NFT collection, creators employ a "smart contract." This script contains digital agreements and policies specific to the NFT project and adheres to open standards like ERC-721 \cite{eip721}. With a shared contract address, each NFT in the collection also receives a unique "token address" for granular asset information. The process of creating and publishing an NFT collection involves uploading the smart contract to a blockchain platform and defining project characteristics like name and total supply. Creators assign a token address to each digital asset which records the asset on the blockchain. Minting typically incurs a gas fee paid by the creator. Following minting, creators list the NFTs for sale on online marketplaces like OpenSea \cite{opensea} or their own websites. When an NFT is purchased, it is transferred to the buyer's decentralized wallet, such as MetaMask \cite{metamask}, and the transaction is recorded on the blockchain.

\subsubsection{Social media promotions:} In addition to targeted advertisements on social media ~\cite{knoll2016advertising}, organizations and brands frequently leverage the help of popular social media individuals (such as influencers) to promote their content in exchange for some financial benefit. Such promotions often involve \textit{sweepstakes}. Sweepstakes are competitions where users have to perform certain activities (such as following the promoted organization and liking and sharing their posts) to participate, after which a winner is drawn at random. Most social media platforms, such as Twitter, have policies to regulate such promotions, requiring the promoter to disclose any compensation for tweeting about a product or service, and prohibiting the use of misleading or deceptive tactics to promote products or services~\cite{twitter_contest_rules}. 
Given the NFT ecosystem's volatility, an NFT collection's success hinges on its popularity/virality, prompting creators to utilize promotions to drive project engagement. In this work, we specifically look at the characteristics and impact of promoting fraudulent NFT projects on Twitter.


\section{Methodology}
\sayak{Change: Introduces the methods for all analyses, instead of having them in their own section in previous submission}

\subsection{Anatomy of Promotion Tweets}
\label{anatomy-nft-promotion}
\begin{figure}[t]
\centering
\includegraphics[width=0.7\columnwidth]{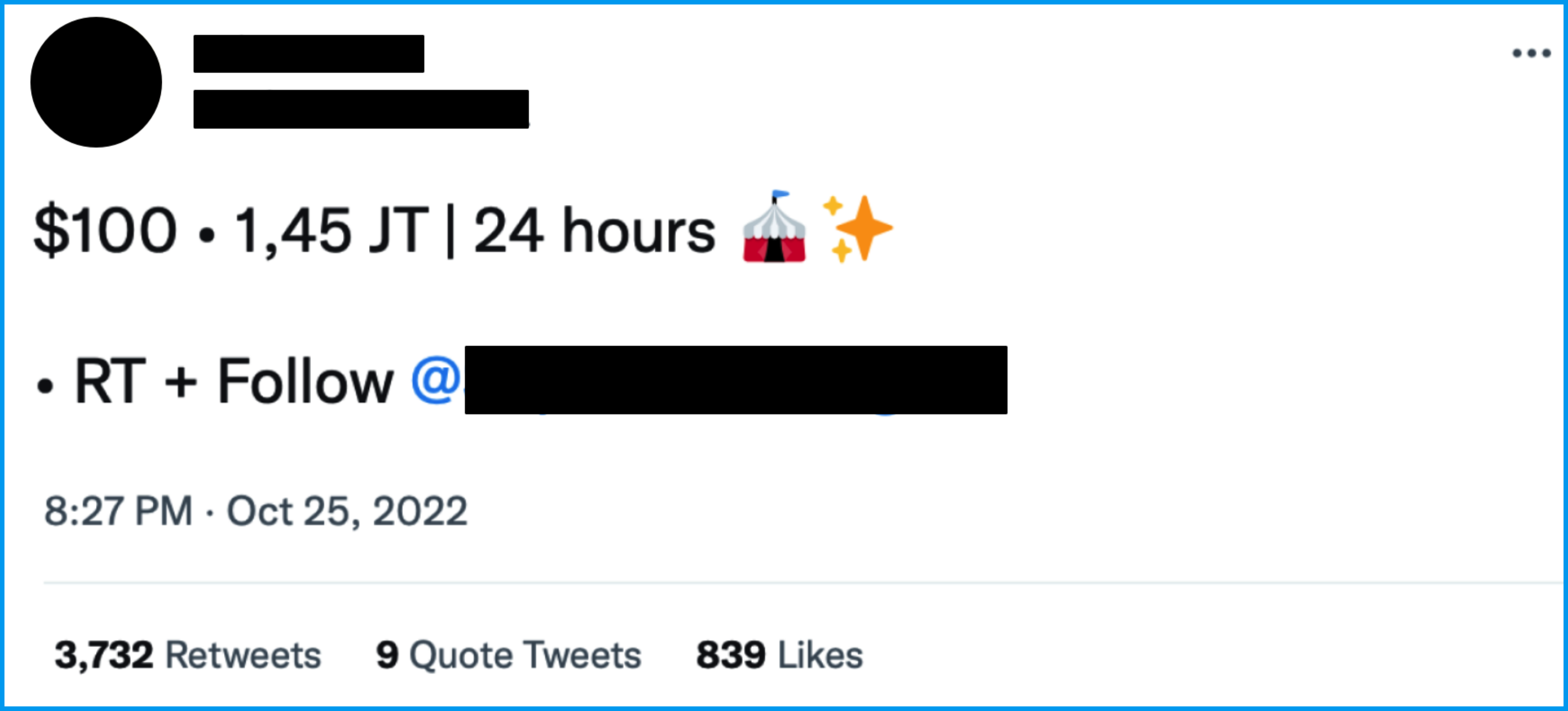}
\caption{Example of a promotion tweet shared by an NFT promoter account advertising a sweepstake competition for a period of one day. Users who retweet this tweet and follow the tagged NFT project have a chance to win $1.45J$ (million) worth of a cryptocurrency token worth \$100}
\sayak{Change: Users anonymized}
\label{fig:tweet_promotion}
\end{figure}

For brevity, in this paper, we refer to  (Twitter) accounts that share promotional tweets for NFT projects as \textit{promoter} or \textit{NFT promoter}, the account of the NFT project that is being promoted as \textit{promotee}, and users who interact with the promotional tweet as \textit{participants}. The tweets shared by \textit{promoters} are typically sweepstake competitions, where the participants have a chance to win money if they follow the account (belonging to \textit{promotee}) and retweet the promotion tweet within a stipulated period of time (ranging from a few hours to a couple of days). 
Incentivizing \textit{participants} to follow the \textit{promotee} artificially increases the followers of the latter, whereas retweeting the promotion tweet further drives traffic to it and might cascade into the followers of the \textit{participants} also participating in the giveaway. 
After the conclusion of the giveaway, the \textit{promoter} randomly chooses one of the \textit{participants} as the winner. The winner can receive the fund directly into their cryptocurrency wallets, or in some cases, in their digital currency wallets. 
Figure~\ref{fig:tweet_promotion} shows an example of an NFT promotion tweet on Twitter. We hypothesize that the artificial inflation of engagement for the NFT collections might paint a false picture that the collection is very popular, which would lure novice buyers into investing. 

\subsection{Collecting NFT Promotion Accounts and Tweets}
\label{collecting-nft-tweets}
To effectively gather data on NFT Promotion accounts and tweets endorsing fraudulent NFT projects, our first step involved recognizing recurring patterns that could aid us in pinpointing these instances. For this task, we enlisted the help of two coders who were well-versed in the cryptocurrency and NFT space. They browsed through Twitter to find 100 tweets that were promoting NFT projects. Among these, a remarkable 94\% of the tweets followed a specific template, as demonstrated in Figure~\ref{fig:tweet_promotion}.
In this sample, 41 unique accounts shared these promotional tweets. Notably, 39 of these accounts utilized the keyword \texttt{NFT promoter} within their profile description. The remaining two profiles were outliers, with one lacking a profile description entirely and the other using Bitcoin emojis instead. The usage of the term \texttt{NFT promoter} suggests self-identification by these accounts, possibly aiming to attract potential customers for their services.
Thus, to find NFT promoter accounts automatically, we collected profiles (using the Twitter API~\cite{TwitterAPI}) that used this keyword in their profile description. Our focus was primarily on accounts with more than 40k followers, as accounts with a large number of followers have more visibility on the platform~\cite{nilizadeh2016twitter} and we also found that they shared promotion tweets more frequently.
By focusing on such accounts, we ensured that our analysis covered promotional tweets that reached a wide audience, thereby increasing the overall impact and relevance of our study.

From June 15 to August 20, 2022, we gathered data from 439 unique accounts fitting our criteria. These accounts collectively shared 21.6k \textit{promotion} tweets. 
To collect these promotional tweets, a regular expression was implemented that matched the tweet template in Figure~\ref{fig:tweet_promotion}. 
Detecting NFT-related promotional tweets presented a distinct challenge, given that this tweet template was frequently used for promoting non-NFT content as well, such as business organizations, products and celebrities. Thus, the two manual coders looked through each tweet to identify if it was an NFT promotion. Our final dataset comprised 2,831 tweets that were shared by 439 NFT promoters which collectively promoted 823 unique NFT projects.

\subsection{Identifying Fraudulent NFT Projects}
While artificial inflation of followers for a less popular (but legitimate) NFT collection might encourage buyers to invest in a low-yield NFT asset, promoting NFT collections that are actually scams might lead users into fraudulent transactions and give away their sensitive information. 
To identify such fraudulent collections for evaluating \textbf{RQ1}, we first manually investigate each unique NFT collection account, focusing heavily on those which had become inactive, i.e., were removed or had been suspended through the course of our analysis using the Twitter API. 
Prior literature has established that Twitter accounts are usually removed by individuals to conceal malicious activity~\cite{volkova2016account}. 
If such malicious activity, however, is caught beforehand, Twitter suspends the account~\cite{twitter_suspended_accounts}.
We look for two characteristics for labeling an NFT collection as fraudulent: i) Accounts that imitate a popular NFT collection and share a fake NFT minting phishing link with the sole goal of stealing their cryptocurrency wallet credentials, and ii) Accounts that had completed their minting period, but had removed or abandoned their website and/or NFT marketplace page, indicating a \textit{rugpull}.   

\subsection{Tracking Engagement}
Followers, retweets, and replies to tweets are associated with online visibility and relevancy of a user~\cite{nilizadeh2016twitter}.
Users often consider content from popular accounts to be reliable and trustworthy~\cite{morales2014efficiency}, which, in turn, makes them even more popular. 
Usually, these metrics increase organically when the user consistently shares the content of substantial interest or value for a wide network of users.
However, in our case, owners of new NFT collections utilize promotions to \textit{artificially} increase their online visibility to drive interest, which can, in turn, can entice users into purchasing the NFT(s). In the case of fraudulent NFT collections, this can lead to victim users being scammed. Thus, to answer \textbf{RQ1}, we find the extent to which promotion tweets increase engagement with NFT projects, especially those that were fraudulent.
First, we track the followers gained by NFT collections through promotion and compare this gain between legitimate and fraudulent collections. Now, bot accounts have been known to play a major role in influencing the narrative of content in social media by engaging heavily with accounts concerned with various political propaganda or misinformation ~\cite{wang2020short}. Thus to accurately identify the impact of promotion in increasing engagement from real users, we investigate the prevalence of bots in engaging with both the promotion and the promoted NFT collections. Specifically, we compare the contribution of bots versus real users in i) Retweeting the promotion tweet, ii) Following the promoted NFT collection \textit{during} and \textit{after} the promotion period, and iii) Interacting with tweets (i.e. likes and replies) posted by the promoted NFT collection \textit{after} promotion. To identify bots, we used Botometer~\cite{botometer}, an automated tool that assigns a score to a Twitter account based on its likelihood of being a bot by evaluating its account characteristics and interactions with other users.  


In conclusion, our research methodology demonstrates a novel, systematic approach  towards collecting NFT promotion accounts and the tweets they share, specifically those endorsing fraudulent NFT projects. This paves the way for the characterization and analysis of these scams throughout this paper.


\subsection{Tracking Anti-Scam Effectiveness}
Prior literature has established that users are highly susceptible to social engineering attacks, which in turn makes it critical for anti-scam measures such as blocklists and browser protection tools, as well as the domain provider (that hosts the website) to identify these attacks quickly and effectively~\cite{oest2020phishtime}. 
For each NFT project in our dataset that was identified as fraudulent, we checked if they were detected by each of four blocklisting entities - Google Safe Browsing, APWG, PhishTank, and OpenPhish at intervals of every $10$ minutes. On the other hand, to determine the number of browser protection tools that detected the NFT-based phishing attacks over the period of a week, we used the VirusTotal API~\cite{VirusTotal:2020}, an online tool that aggregates the detection rate of 76 such tools. 
Prior literature~\cite{peng2019opening} has explored the possibility of VirusTotal labels lagging behind their respective anti-phishing tool engine, and to negate this effect, we scanned each URL regularly at intervals of 10 minutes throughout the study period. Finally, we checked if the NFT phishing websites were active by sending HTTP GET requests to them every $10$ minutes (with a response code of 200 being considered as being active). Since more resilient phishing attacks can evade this approach~\cite{oest2020phishtime}, we also monitored if the screenshot and code base of the website changed significantly to indicate that it had become inactive. These analyses collectively aid us in answering \textbf{RQ2}.

\subsection{Tracking Sales of Promoted Collection} 
The surge in the popularity of  \textit{promoted} fraudulent NFT projects, driven by an artificial increase of followers and retweets due to the \textit{promotion} itself, might encourage unsuspecting users to invest in such scams~\cite{kapoor2022tweetboost}.  
Thus, to answer \textbf{RQ3}, we determine the volume of monetary transactions made towards these scams. 
We collect the contract addresses for each NFT project, and use Etherscan~\cite{etherscan} and BscScan~\cite{bscscan}, which are open-source blockchain explorers, to  analyze transactions made to the cryptocurrency wallets of the attackers.


\section{Characterization and Detection of Promoted NFT Projects}
\label{results-nft-promotion}
\sayak{Change: All analyses in Section 4 and 5 in the initial submission is combined in this section}
\subsection{Categorizing Promoted NFT Scams}
\label{categories}
Through manual analysis of all 823 promoted NFT projects, we were able to find 300 fraudulent NFT projects that fell into one of three broad categories:
\subsubsection{Phishing Scams:}
\label{phishing-scams}
We found $22.1$\% (n=$182$) of the promoted NFT projects were imitating a legitimate NFT collection, and sharing phishing links to steal the victim's cryptocurrency wallet credentials. These scams contain a minting link for users to purchase one or more tokens that are in ``\textit{high demand}.'' Upon clicking the link, users are prompted to provide full transaction rights to their cryptocurrency wallet, such as Metamask~\cite{metamask}. This action enables the attacker to transfer assets, including funds and NFT tokens, from the victim's wallet. Figure~\ref{fig:nft_phish_example} illustrates an example of an NFT phishing attack imitating the Primates Embassy NFT collection. We were able to extract the NFT contract addresses of 57 out of these 182 phishing websites, which we use to further characterize NFT based phishing scams in Section~\ref{nft-phishing-scams}. 

\subsubsection{Pre-mint Scams:} We also found $14.4$\% (n=$119$) promoted NFT projects that never went to the minting phase. Developers usually announce their minting date months in advance, and we consider these collections to be abandoned \textit{iff} it was already past their minting date, and the developer had not posted a new tweet in two months. While the notion of pre-mint projects does not indicate malicious activity, we found $40$ collections that imitated a popular NFT project and had set up a pre-mint website where they ask users to connect their cryptocurrency with \textit{full transaction} rights such that they can obtain \textit{whitelist} spots that give users priority access to the NFTs when it begins minting (which never happened). While not malicious at first glance, asking for full transaction rights to the user's cryptocurrency wallet is a notable characteristic of NFT phishing.

\subsubsection{Rugpull Scams:} Finally, we discovered that 9.4\% (n=78) of the promoted NFT projects exhibited characteristics of rug pulls. NFT rug pulls are projects that are often launched with an appealing website, engaging social media presence, and enticing artwork. They successfully attract a substantial number of buyers before suddenly disappearing or becoming inactive. We identify projects as rug pulls when they had at least two of the following characteristics: i) The project already held an NFT minting stage, but their contract address shows no transfer of assets; ii) The author of the project has not provided any updates for the last two months, usually after completing the minting stage, and iii) User feedback on the author's profile indicates that users did not receive the NFT assets after the minting stage.

\subsection{Tracking Visibility and Engagement of Promoted NFT Projects}
\subsubsection{Removed and Suspended Accounts:}More than $18.1$\% (n=$149$) of the promoted NFT project accounts were removed, whereas $8.9$\% (n=$74$) accounts were suspended by Twitter. The median time of removal and suspension of these accounts was 231 days and 157 days respectively. As we will see in the next section, the majority of these accounts comprised of NFT scams, and this is a substantial amount of time for malicious actors to carry out the attack, especially considering that other online scams usually last for less than a week~\cite{atkins2013study}.

We found that 61.1\% (n=91) of the projects whose Twitter accounts were removed and  64.7\% (n=48) of projects whose accounts were suspended were imitating a legitimate NFT collection, and sharing phishing links to steal the victim's cryptocurrency wallet credentials. 
Also, $29$ NFT collections sharing phishing links were active during the entire duration of the study.  
On the other hand, 34.8\% (n=52) of the removed collection Twitter accounts, and 16 (21.6\%) of the suspended collections Twitter accounts, after having completed their minting phase (i.e., when users can purchase the NFT tokens in exchange for cryptocurrency funds), had abandoned the project, indicating a rug-pull. 



\begin{table}[]
\centering
\resizebox{0.4\textwidth}{!}{%
\begin{tabular}{c|c|c|c|c}
\hline 
Status & Min & Max & Mean & Median \\ \hline \hline
Active & 2 & 37,087 & 4,195.98 & 2,601 \\ 
Removed & 2 & 324,091 & 5,412.85 & 2,159 \\ 
Suspended & 2 & 55,522 & 3,203.59 & 1,265 \\ \hline
\end{tabular}}
\caption{Descriptive statistics of followers gained by NFT collections due to promotion}
\label{descriptive-stats-promotees}
\end{table}

\begin{figure}
  \centering
  \includegraphics[width=0.4\textwidth]{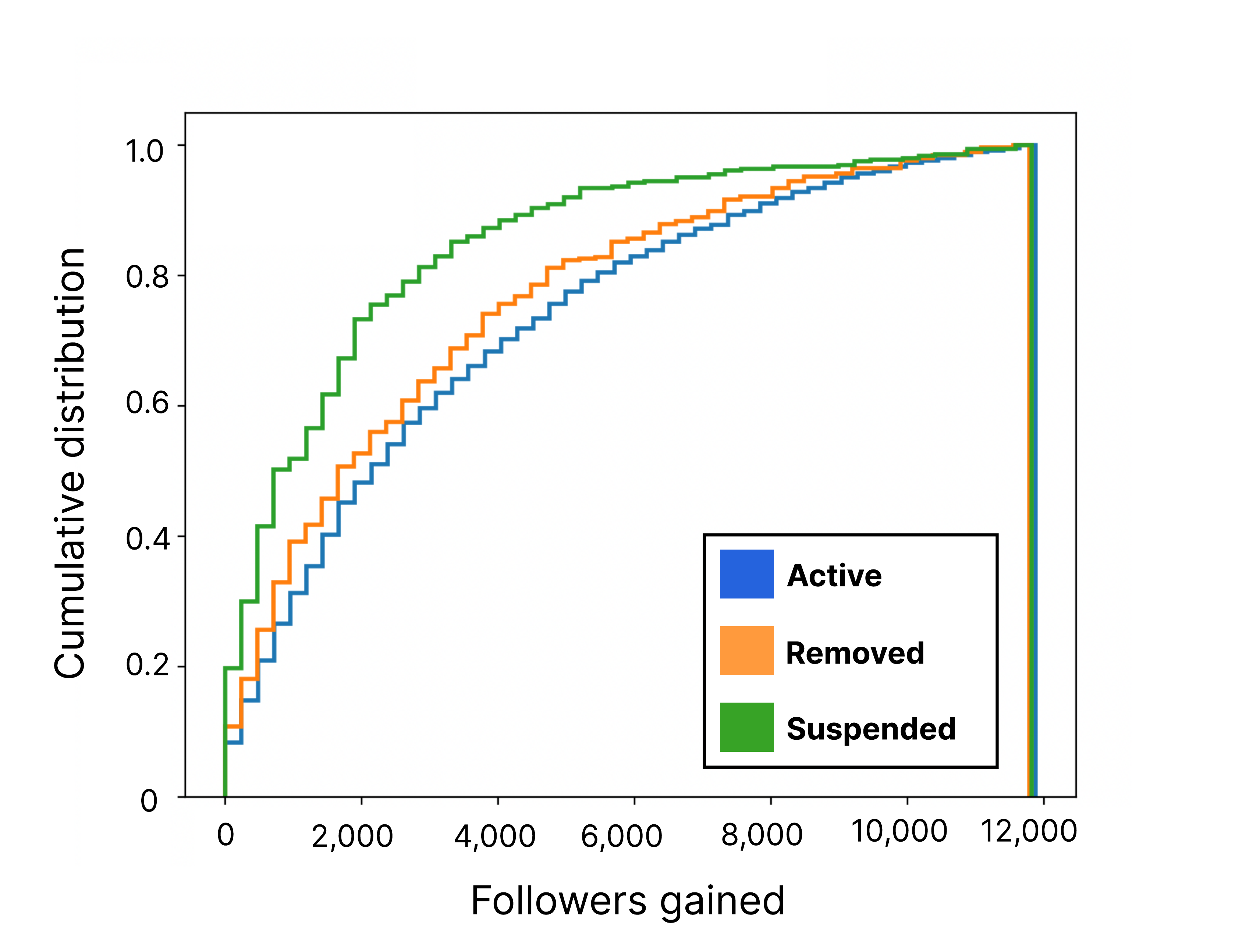}
  \caption{Cumulative distribution of followers gained by Active, Removed, and Suspended NFT accounts that were promoted.}

  \label{fig:follower-gain-cdf}
\end{figure}
\subsubsection{Increase in Followers during Promotion:}
\label{promotion-followers-engagement}
We identify the increase of followers of the fraudulent NFT projects that are promoted on Twitter.
Considering that the vast majority ($92.8$\%) of accounts that were removed or suspended turned out to be fraudulent, we compare the increase in followers for these accounts versus those that were active throughout our study. We find that NFT collections promoted through Twitter gain a substantial number of followers over their promotion period, as indicated by their descriptive statistics illustrated in Table~\ref{descriptive-stats-promotees}. By constructing a cumulative distribution function of followers gained by each promoted NFT collection, as illustrated in Figure~\ref{fig:follower-gain-cdf}, we found that the followers gained by \textit{removed} accounts were very similar to those that were active, while accounts that were suspended also had a significant rise in followers. This suggests that removed and suspended accounts (that hosted NFT scams) were just as likely to gain a large number of followers as their legitimate counterparts through promotion. This is problematic, as it can encourage attackers to use these (promotion) services.  

\subsubsection{Bot Participation and Organic Engagement:} 
\label{bots-organic-engagement-promotion}
\begin{figure*}
    \centering
        \subfloat[Retweets]{\includegraphics[width=0.27\textwidth]{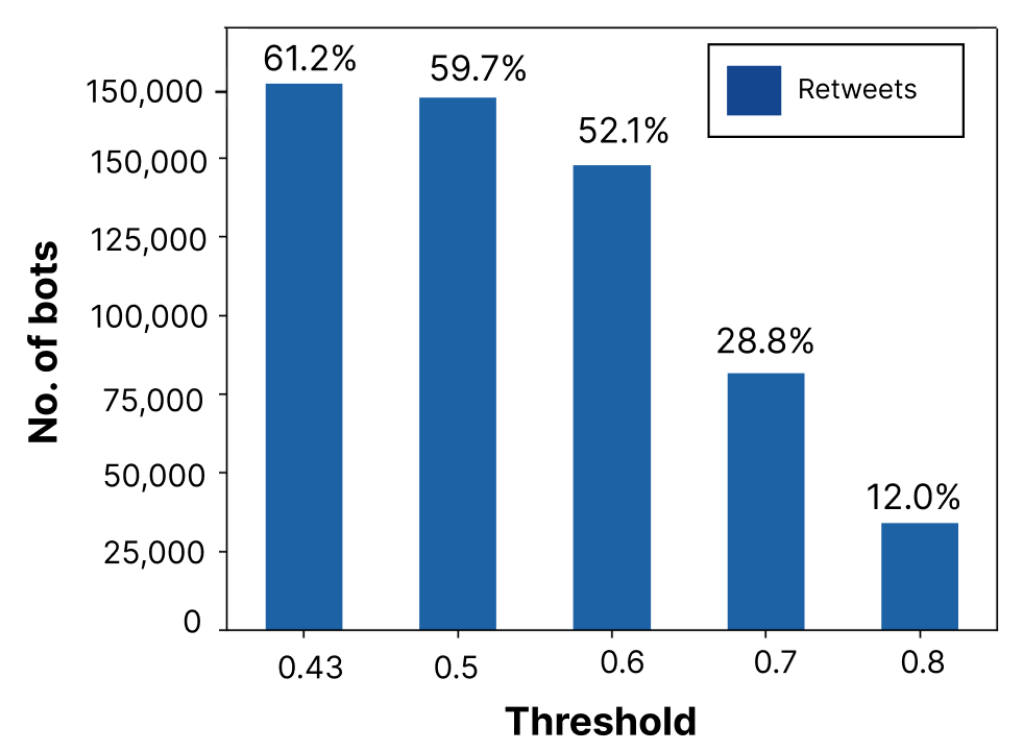}\label{fig:threshold-retweets}}
        \subfloat[Followers before/after]{ \includegraphics[width=0.26\textwidth]{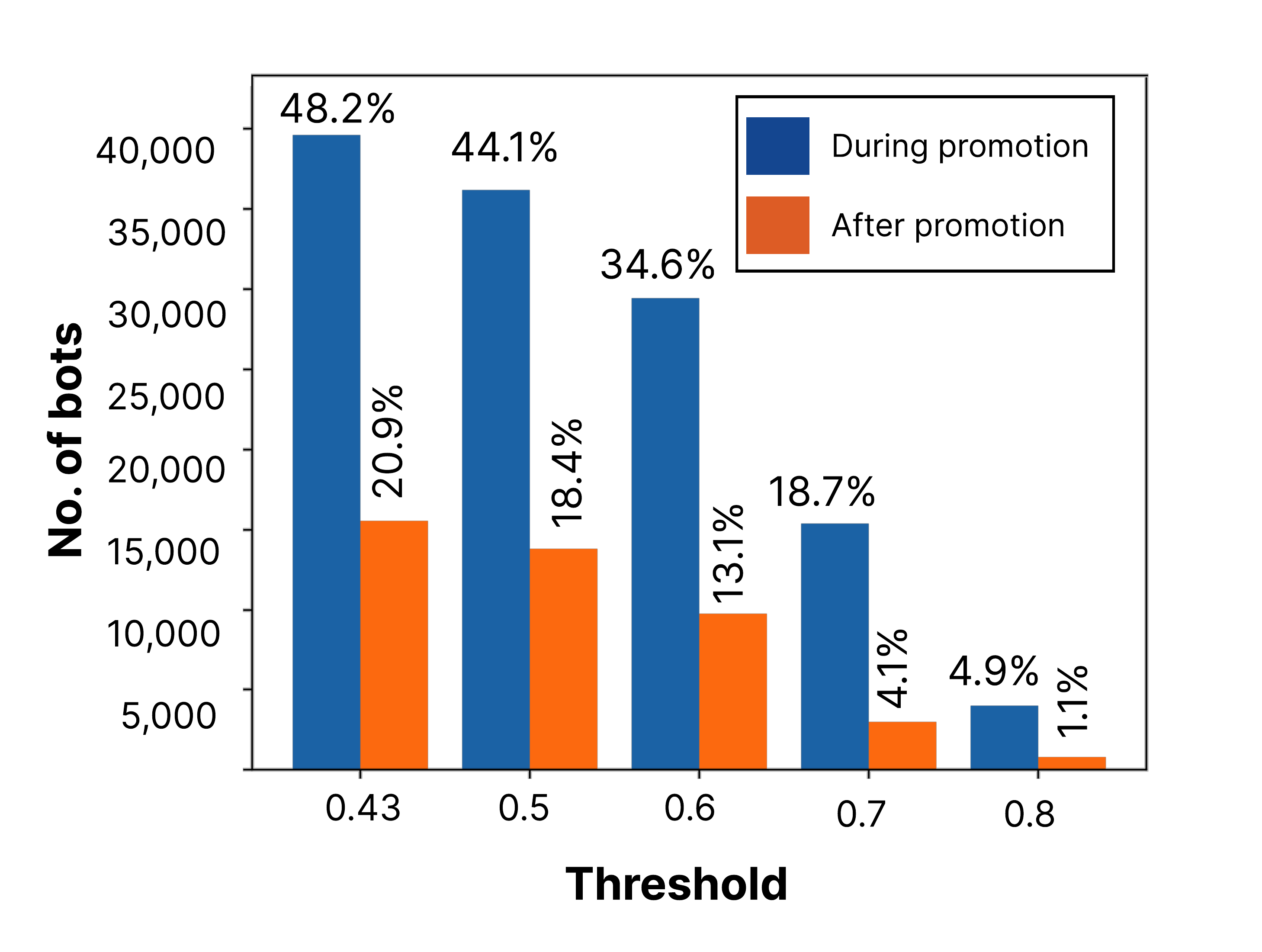}\label{fig:threshold-follows}}
        \subfloat[Likes/Retweets]{ \includegraphics[width=0.26\textwidth]{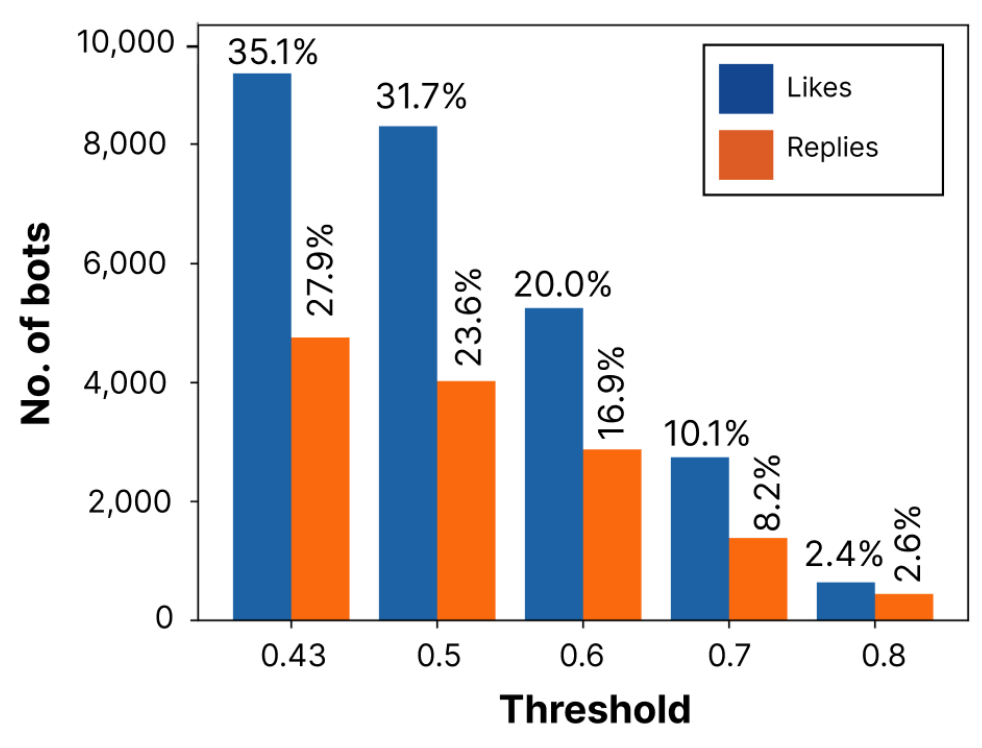}\label{fig:threshold-likes-replies}}
       \caption{Distribution of bots across different Botometer score thresholds for \textbf{(a)} Bot accounts who had retweeted the promotion tweets,\textbf{ (b)} Bots following NFT collections during and after promotion and \textbf{(c)} Likes and replies by bots on tweets shared by NFT collections.}
  \label{fig:bot-thresholds}
\end{figure*}

We also determine how bots influence  engagement towards the fraudulent NFT collections that were promoted. Since the participants are asked to retweet the promotion tweets, we randomly selected 100 users who had retweeted each promotion tweet, collecting the impressions of $283,409$ total users in the process. For our study, we used the Botometer scoring threshold $t = 0.43$, which is in line with several prior works relevant to this domain. ~\cite{pew_research_bots_twittersphere,shao2018spread} However, the optimum threshold has been up a subject of much debate~\cite{yang2022botometer}. Thus, we provide a histogram for bots across different scoring thresholds for each of our experiments in Figure~\ref{fig:bot-thresholds}.
While the number of bots decreases with the increase of $t$, we notice a similar pattern of disparity between bots and organic user engagement across all thresholds and thus interpret our analysis using $t =0.43$.

We found that more than 61\% of users (n=$173,446$) who had retweeted promotion tweets exhibited bot-like activity. The distribution of bots across other score thresholds is illustrated in Figure~\ref{fig:threshold-retweets}.
To check if bots also started following the promoted accounts,  we randomly picked 100 new followers both \textit{during} the promotion period (n=$82,139$), and a week \textit{after} the promotion ended (n=$74,018$). 
For NFT collections that were promoted multiple times, the last \textit{promotion} period was considered for this analysis. 
We found more than $48.2$\% of the new followers \textit{during} the promotion were perceived as bots, whereas the statistic was much lower at $21.7$\% post-promotion. 
To also investigate this issue on a per-user basis, we conducted a Wilcoxon Rank Sum Test and found that \textit{promotees} were more likely to gain followers which were bots \textit{during} the promotion period than \textit{after} (p\textless$0.01$). The distribution of bots for this experiment across other score thresholds is illustrated in Figure~\ref{fig:threshold-follows}.
Next, to identify if real users were interacting with the NFT collection after promotion, we consider (at most) the first five tweets shared by the \textit{promotee} \textit{after} at the end of the promotion, and randomly select $100$ users who had liked each tweet and also a maximum of 10 replies.
We evaluated $3,897$ such tweets, which  trggered 26,182 likes and 17,039 comments. We found $64.9$\% of all likes came from real accounts, with the remaining ones being bots, 
whereas $72.1$\% of all comments came from real accounts, with the rest being bots. We also note the distribution of bots for this experiment across other score thresholds in Figure~\ref{fig:threshold-likes-replies}.
While most of this engagement is organic, we assume that the engagement from bot accounts might be due to these projects also promoting their (fraudulent) NFT links ~\cite{facebook_twitter_nfts}.
 
To summarize, we observed that bots significantly participate in promotions by increasing the follower count of the fraudulent NFT collections and also retweeting their tweets during the promotion period. This leads to real users following and engaging with the \textit{promoted} fraudulent NFT collection later on, which eventually leads to them transferring money to these scams, which we discuss in Section~\ref{financial-impact-promoted-tweets}. 

\subsubsection{Suspensions and Modifications of Promoters:}
\label{suspension-modifications-promoted-accounts} 
We found $46.2$\% of the NFT promotion accounts (n=$203$) had been suspended throughout the course of our analysis. 
These accounts collectively promoted nearly $54$\% (n=$129$) of all fraudulent NFT collections in our dataset, and the age of these accounts was significantly lower (Age\textsubscript{median}=$191$ days) compared to the age of the accounts which were not suspended((Age\textsubscript{median}=$478$ days, p\textless$0.05$\textsuperscript{*}).
We also found $12.1$\% (n=$53$) of the accounts which were not suspended had deleted at least half of their tweets since the start of their analysis.
This behavior was also seen in $9.4$\%(n=$83$) of promoted NFT collections. 
Since these accounts could have also modified their account information, we did not have the necessary data (i.e., minting website/contract address) to verify if they were fraudulent, and thus conservatively marked them as \textit{legitimate} for our analysis.  
For both promoter and promotee accounts who had removed their previous tweets, we assume that they hide their previous activities to evade detection by Twitter's anti-spam measure~\cite{twitter_platform_manipulation}, and retain the large number of followers they already have to promote/host another scam later on. 
\subsection{Characterizing NFT Phishing Scams}
\label{nft-phishing-scams}
Rugpull scams are nearly identical to legitimate NFT projects right up to the minting (sale) stage, thus (unfortunately) containing little to no features that can be utilized to proactively identify them. However, this is not the case for phishing scams and fradulent pre-mint projects that we manually evaluated in Section~\ref{phishing-scams}. NFT Phishing scams imitate a legitimate NFT project and entices potential victims into \textit{purchasing} a token that might already be \textit{sold out} or in \textit{high demand}, and while NFT pre-mint scams entice users to connect their cryptocurrency wallet to participate in the minting process at a future date.
Through our manual analysis of these fradulent projects, we have identified two primary attack vectors used in these scams:
\begin{figure}[h!]
\centering
\includegraphics[width=0.6\columnwidth]{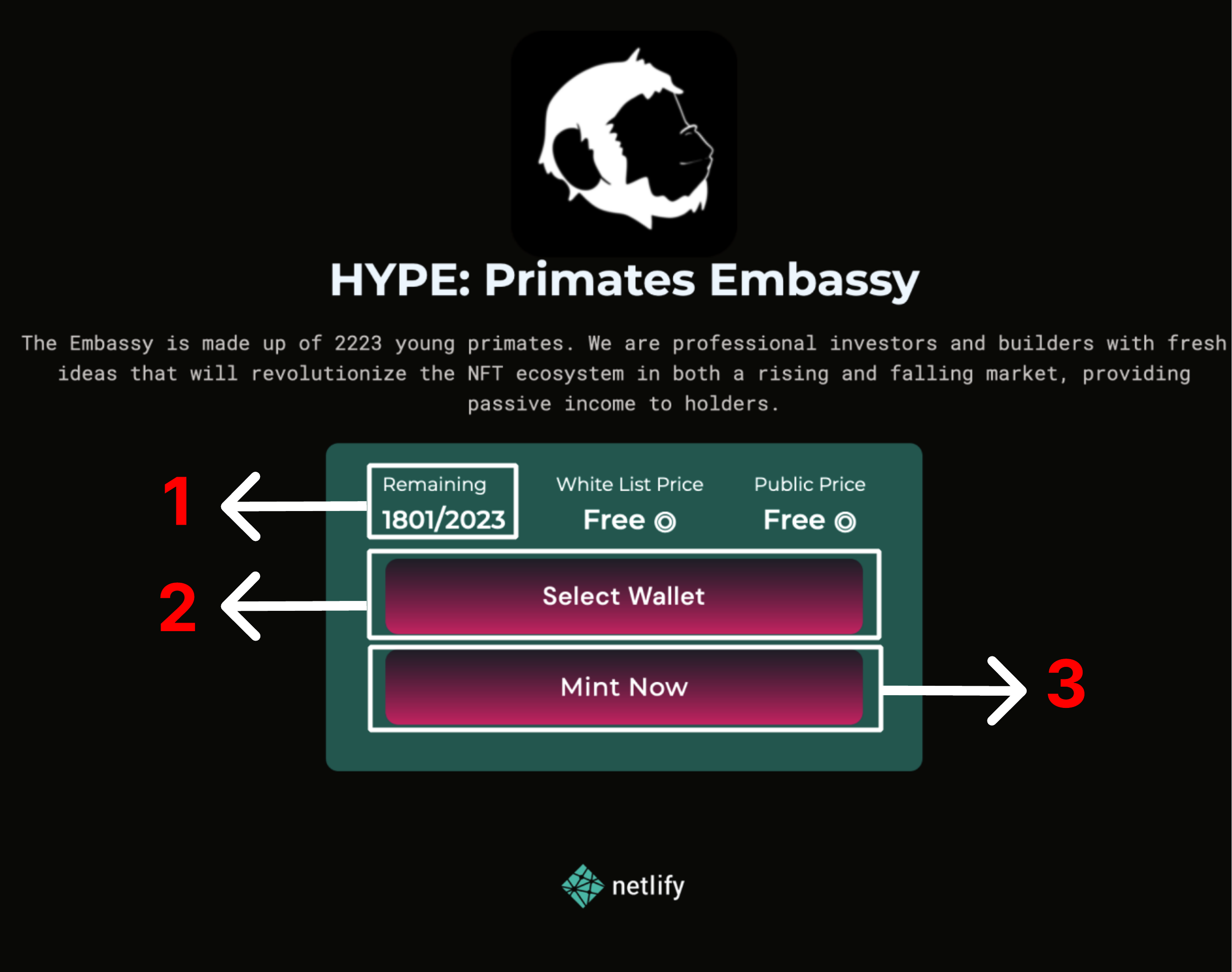}
\caption{An example of an NFT phishing attack imitating the Primates Embassy NFT collection. Feature \textbf{(1)} demonstrates an artificial counter, \textbf{(2)} represents the victim connecting their wallet and granting full transaction rights, and \textbf{(3)} shows the execution of the malicious payload, which steals funds and/or NFT tokens from the victim's wallet.}
\label{fig:nft_phish_example}
\end{figure}
\subsubsection{Fraudulent Fund Transfers:}
\label{phish-fund}
For 41 phishing URLs, we found that, upon connecting the cryptocurrency wallet, the website initiates the transfer of a specified amount to the attacker's wallet. Given that the attacker has full access to the victim's wallet, they can also initiate transactions at a later time, even if the victim is no longer active on the website.

\subsubsection{NFT Token Theft:}
\label{phish-token-transfer}
We discovered 16 phishing URLs in our dataset containing several contract addresses embedded within the website source code that belong to legitimate NFT tokens. After obtaining full transaction rights to the victim's wallet, the website checks for a list of popular NFT tokens. If any are found, the tokens are transferred to the attacker's wallet. Figure~\ref{fig:nft_steal_function} shows an example of a code snippet employed in such attacks. The attacker uses the \texttt{syncNfts()} function to identify all NFT tokens held by the user. Upon the user clicking the ``Mint'' button, the tokens are transferred to the attacker's wallet using the \texttt{Send} function.
Popular NFTs often have a high value, sometimes reaching thousands of dollars~\cite{opensea_rankings}. The decentralized nature of NFT transactions renders stolen assets almost impossible to recover~\cite{wu2021analysis}, making this attack vector particularly dangerous.

\begin{figure}[h!]
\begin{lstlisting}[
    language=JavaScript,
    backgroundcolor=\color{lightgray},
    frame=single,
    caption=,
    stepnumber=1,
    firstnumber=1,
    basicstyle=\ttfamily \scriptsize]
async function claim() {
  let user = Moralis.User.current();
  if (!user) {
    await connect(); await syncChain(); await syncNfts();
    btn1.html("MUTATE KODA"); btn2.html("CLAIM");
    await send(); return;
}
\end{lstlisting}
    \caption{Example of a malicious function that steals NFT tokens from the victim's wallet.}
    \label{fig:nft_steal_function}
\end{figure}

\subsection{Distinctive Characteristics of NFT Phishing Attacks} 
\label{evasive-nft-phish}
\shirin{we should not call them evasive features. A feature is evasive if it is intentionally added to something to evade detection. Here we should just say this type of phishing is different from previous ones.}\sayak{Changed to Distinctive Characteristics of NFT Phishing Attacks }
\shirin{if you call it a qualitiative analysis, then you should mention the details of this analysis. How did you do it?}\sayak{Expanded below}
We conducted an in-depth analysis of each of the websites belonging to the fraudulent NFT projects by examining their unique features, tactics, and characteristics. 
This analysis included reviewing the website's user interface, their source code to identify the communication methods used by the attackers and the type of information solicited, and the use of URL-based features. We then categorized these features into common themes and patterns. This step involved iterative coding and recoding of the data until we identified three primary features that make NFT phishing attacks characteristically different from regular phishing attacks.
\shirin{this section can be shorten. I did for Lack of credential-requiring fields. Please shorten the rest.} \sayak{Done}
\textbf{i) Lack of credential-requiring fields:} A primary feature of regular phishing attacks is that they contain text fields that ask for sensitive credential information from potential victims, such as account passwords, credit card information, Social Security number, etc. 
However, we did not observe any NFT phishing attack with such fields. 
\textbf{ii) Using cost-effective TLDs:} 
Regular phishing attacks often purchase cheaper TLDs (such as .xyz, .club, .store, etc.) in bulk to extend their volume and lifespan~\cite{oest2020sunrise}. This practice attracts increased scrutiny from anti-phishing entities who associate these TLDs with phishing attempts. Interestingly, this trend is also common in legitimate NFT projects run by hobbyists, independent artists, or small businesses who cannot afford pricier .com domains. Consequently, these legitimate NFT entities also use economical TLDs, adding complexity to the detection process.
\textbf{iii) Lack of brand identifiability:}
Conventional anti-phishing tools utilize comprehensive databases of typical targets for phishing attacks. For instance, a site like \textit{http://chase-login.com}, soliciting Chase banking credentials, is readily identifiable as a potential phishing site. In contrast, the ever-expanding and rapidly evolving landscape of NFT collections complicates the process of maintaining an updated list of potential targets for NFT phishing attacks, rendering traditional identification strategies less effective.
\subsection{Automatic Detection of NFT-Based Scams}
\label{ml-model-nft-phish}
Through manual evaluation of the websites, we identified key features that enabled us to develop an ML-based detection model to proactively identify these attacks. 
\subsubsection{Building the Ground Truth:}
\sayak{Change: URLs collected from DNSTwist/Certstream are only used for training the model, and not for blocklist analysis}
Our promotion tweets dataset contained only 57 true positive phishing URLs, which were insufficient for creating a ground truth for ML model training. Thus, we collected and verified $1,028$ more unique NFT phishing URLs in two seven-day batches: July 8th, 2022 \textit{(Batch~1)} and October 15th, 2022 \textit{(Batch~2)}. We employed two approaches:
\textbf{(i)} Utilizing DNSTwist, a DNS fuzzing framework, we discovered URLs that used typo-squatting and domain-squatting techniques to mimic the top $100$ NFT collections on OpenSea (by sales volume).
\textbf{(ii)} By applying NFT-specific perturbation terms like "nft", "claim", and "mint", we identified newly registered domains associated with NFTs using the Certificate Transparency Log network~\cite{certstream}. 
For each of the URLs in our dataset, we gathered full website snapshots, including codebases and screenshots. The websites were also manually evaluated by the authors to avoid any \textit{false positives}. Interestingly, all of these URLs utilized either the fund-stealing or token-stealing attack vectors, signifying the robustness of our characterization, despite being done on a small dataset. Additionally, we added $1,471$ benign NFT minting URLs to our ground truth dataset, collected our initial dataset of promotion tweets and OpenSea. Similar to the phishing URLs, these websites were also manually evaluated to avoid \textit{false negatives}. Thus our final ground truth dataset for training the ML classifier contained a total of $1,085$ true positive phishing URLs (i.e. $57$ from our initial observation, and $1,028$ from Cestream/DNSTwist) and $1,471$ benign URL samples. 

\subsubsection{Feature Extraction:} 
\label{ml_features}
We extract the following features from each of the  websites in our ground-truth dataset to train our classification model: \textbf{1.} If the URL matches any of the known NFT collections. \textbf{2.} If the contract address used by the website for the transaction matches that of any known NFT collection. \textbf{3.} The number of Ethereum addresses found in the source code, because attacks with the motive of stealing NFT tokens embed many popular NFT tokens in the website. \textbf{4.} Checking for Twitter links shared by the website. Unlike legitimate minting pages, several NFT phishing attacks do not include links to Twitter pages. \textbf{5.} As seen in our promotion analysis, even if these websites share Twitter links, they might get suspended or removed. Thus we also check if the linked Twitter account is active. 
\textbf{6.} Checking if the Twitter page belongs to a known collection. \textbf{7.} Check if the Opensea page belongs to a known collection. \textbf{8.} Checking for the number of Twitter followers. While we see that fraudulent NFT collections utilize promotions to get followers, that number is still far lower than popular and established NFT collections that they imitate. \textbf{9.} Checking for the age of the Twitter account, as fraudulent accounts tend to have a short lifespan~\cite{thomas2011suspended}, 
\textbf{10.} Checking if the name of the collection attached to the contract address is identifiable using EtherScan~\cite{etherscan}. Older contract addresses belonging to legitimate NFT collections are usually identifiable. By \textit{known} NFT collections, we refer to the top 1K Ethereum-based NFT collections on OpenSea~\cite{opensea}.


\subsubsection{Model Training and Performance} 


We trained our dataset using four algorithms: Decision Tree, Logistic Regression, Support Vector Machine (SVM), and Random Forest (RF). Out of all the models, RF performed the best. To avoid over-fitting, we ran our model through a 10-fold cross-validation study with a training/test set split of 70:30, with $1,789$ websites used for the training set, and $767$ used for the testing set. Our model shows an accuracy of 0.97, with a precision and recall of 0.95 and 0.98, respectively. We also conducted an examination of feature importance within our trained Random Forest model to determine the significance of different features in classification tasks. The top features, along with their respective scores, are: \textit{The Etherscan/BSC scan token was known} (0.218), \textit{The number of Twitter followers} (0.134),\textit{The age of the Twitter account} (0.125), \textit{the reputation of the Minting URL} (0.123), and the \textit{Contract address} (0.121).


\subsection{Comparison with Other Models}
\sayak{New section for R\&R}
\label{comparison-with-other-models}
\begin{table}[]
\caption{Comparison with other classification models}
\label{model-comparison}
\resizebox{0.5\textwidth}{!}{%
\begin{tabular}{c|cccc}
\hline
Model & Accuracy & Precision & Recall & F1-score \\ \hline
VisualPhishNet & 0.34 & 0.29 & 0.24 & 0.26 \\ \hline
URLNet & 0.30 & 0.24 & 0.18 & 0.21 \\ \hline
PhishPedia & 0.52 & 0.48 & 0.41 & 0.44 \\ \hline
StackModel & 0.38 & 0.32 & 0.27 & 0.29 \\ \hline
\textbf{PhishNFT (our model)} & \textbf{0.97} & \textbf{0.95} & \textbf{0.98} & \textbf{0.95} \\ \hline
\end{tabular}}
\end{table}
We compared the performance of our classifier, PhishNFT, with four state-of-the-art ML-based phishing detection models: two that rely on the visual features of the website: VisualPhishNet~\cite{abdelnabi2020visualphishnet} and PhishPedia~\cite{lin2021phishpedia}, one that relies on both the URL string and  HTML representation of the website: StackModel~\cite{li2019stacking}, and one that relies on the semantic representation of the URL string only: URLNet~\cite{le2018urlnet}. All models were tested across the 767 URLs in our testing set. 
Table~\ref{model-comparison} shows that our classifier significantly outperforms all other models on NFT phishing attacks, having a true positive rate (recall) of 0.98 that is nearly \textit{four times}, \textit{three times}, and more than \textit{four times} higher than VisualPhishNet (0.26),  StackModel (0.29) and URLNet (0.21), respectively. PhishPedia had a comparatively better performance than the other baselines with a recall of 0.44.

\subsubsection{Implications:}
VisualPhishNet and PhishPedia primarily rely on visual similarity and credential-collecting fields for detecting phishing websites, which are absent in NFT phishing pages and can thus lead to false negatives.
Additionally, PhishPedia's reliance on brand logo identification may not be effective for NFT projects. 
URLNet and StackModel focus on the structure of the URL for detection. However, as discussed in Section~\ref{evasive-nft-phish}, many legitimate NFT creators use cheaper domains, which are also often used by attackers for hosting phishing pages. This makes it challenging to differentiate between legitimate and fraudulent NFT websites using URL structure alone. StackModel's additional reliance on DOM features of traditional phishing websites may further limit its effectiveness.
\shirin{changed here, please check:} \sayak{Looks good}
On the other hand, our model is tailored to NFT-scam-specific features.
Of course, our approach can be used side-by-side with other phishing detection models that focus on detecting non-NFT phishing attacks. 
\subsection{Identifying New Fraudulent NFT Phishing Projects}
\label{identifying-new-scams}
\sayak{New section for R\&R}.
In the next step, we employed our detection model to identify, in real-time, new fraudulent NFT phishing projects promoted on Twitter.
We ran our system from February 2nd to April 25th, 2023. In total, we discovered 401 new fraudulent NFT projects shared by 87 promotion accounts, 71 of which were not present in our initial dataset. Out of them, we manually verified 382 of them to be true positives, thus resulting in a true positive rate of 95.2\%.
These accounts had a median follower count of 3.7k and a median like count of 198 (based on the first 10 posts in their timeline), indicating noticeable engagement. The fraudulent accounts, after being detected by our tool, were reported to Twitter~\cite{twitter2021reporting}.
We also reported the URLs associated with these accounts to their hosting providers, to aid their removal.
However, we found that only 59 out of 382 fraudulent NFT projects (approximately 15\%) were suspended by Twitter within a week of them being shared by a promotion account, and a collective total of 134 accounts (about 35\%) were suspended during the whole duration (from February 2nd till April 29th). This suggests that Twitter's response to detecting these scams is not only slow, but their overall coverage is also low during the first week. 
It is also notable that 53 out of the 71 new promoters identified by our classifier contained the word ``NFT Promoter'' in their profile description. Tweets posted by these promoters that shared the fraudulent NFT projects had a median like and retweet count of 2.2k and 1.4k respectively - indicating that the phishing projects promoted through these accounts potentially had substantial reach. However, none of these promotional accounts were suspended by Twitter within a week, indicating that these promotion accounts are an unrestricted outlet to promote fraudulent NFT projects.
\section{Evaluating Prevalent Anti-Scam Measures} 
\label{anti-phishing-effectiveness}
\sayak{Change: Not only phishing but all NFT scams are included in this analysis. Language for anti-phishing remove and replaced by anti-scam or browser protection tools}
\sayak{Show performance against rug pulls and premints as well}
Prior work has established that users are susceptible to social engineering attacks.
Thus, it is critical for anti-scam measures, such as blocklists and browser protection tools, as well as domain providers (that host malicious websites) to identify and remediate attacks quickly and effectively~\cite{oest2020phishtime}. 
In this section, we address \textbf{RQ2} by assessing the performance of existing protection mechanisms against NFT-based scams. We run a longitudinal analysis for each URL associated with the 382 fraudulent projects that were found by our classifier in Section~\ref{identifying-new-scams}, as well as the 97 fraudulent projects (57 phishing scams and 40 pre-mint scams) identified during our manual analysis in Sections~\ref{phishing-scams} - for a week from their first appearance on Twitter to determine their \textit{detection rate} and \textit{detection speed} (i.e., how quickly the website was detected/taken down) by popular antiphishing entities. 
We consider four popular blocklists: PhishTank~\cite{phishtank}, OpenPhish~\cite{openphish}, APWG eCrimeX~\cite{ecrimex:2022} and Google Safe Browsing~\cite{gsbdeveloper} \textbf{(Section~\ref{blocklist})}. We also evaluate these attacks against 76 anti-phishing tools using VirusTotal~\cite{VirusTotal:2020} \textbf{(Section~\ref{browser-tool})}, as well as the domain providers hosting the respective websites \textbf{(Section~\ref{domain})}. 
Since these websites are a new family of phishing attacks, we also gauge how their detection rate and speed compare with that of regular phishing attacks for all relevant anti-scam measures in their respective sections.
For our analyses, we compare the NFT phishing URLs that we found manually (n=57) as well as those using our automated detection tool (n=382 true positive URLs). To further contrast the performance between these URLs versus regular phishing attacks, we performed the same evaluation on a random sample of the same number of regular phishing attacks, which we collected from Certstream~\cite{certstream}. We did not include the URLs that we collected separately from DNS records for training our automated model (n=1,028, highlighted in Section~\ref{ml-model-nft-phish}), since we do not have any information on how and where those URLs were distributed. 

\subsection{Blocklist Performance} 
\label{blocklist}
Table~\ref{table-blocklist-performance} shows the summary of the performance of blocklisting entities. PhishTank detected only $5.8\%$ of NFT phishing URLs with a median detection speed of 9 hrs 49 mins, compared to $15.4\%$ of regular phishing URLs at a median speed of $76$ mins. 
The disparity in both \textit{detection rate} and \textit{speed} was also significant across the other three blocklists. 
For example, Google Safe Browsing, the default anti-phishing blocklist in several Chromium-based browsers and Mozilla Firefox, was able to detect $79.9\%$  of all regular phishing URLs at a median detection speed of 84 mins, compared to only $18.8$\% of all NFT phishing attacks at a much slower median speed of 4 hrs 56 mins. On the other hand, \textit{none} of the URLs for the rug-pulled projects (n=31) were detected throughout the duration of the study, and only two of the pre-mint scams (n=29) were detected. 

\begin{table}[]
\caption{Blocklist performance of anti-phishing blocklists against NFT-based phishing attacks found by our initial manual analysis (n=57 and by our automated detection tool (n=382)\sayak{Change: Updated stats with the new 382 URLs}}
\resizebox{0.9\columnwidth}{!}{

\label{table-blocklist-performance}
\begin{tabular}{|l|lc|lc|}
\hline
\multirow{2}{*}{\textbf{Blocklist}} & \multicolumn{2}{c|}{\textbf{NFT-attacks}} & \multicolumn{2}{c|}{\textbf{Regular-attacks}} \\ \cline{2-5} & \multicolumn{1}{c|}{\textbf{Coverage}} & \textbf{\begin{tabular}[c]{@{}c@{}}Median speed\\ (hh:mm)\end{tabular}} & \multicolumn{1}{c|}{\textbf{Coverage}} & \textbf{\begin{tabular}[c]{@{}c@{}}Median speed\\ (hh:mm)\end{tabular}} \\ \hline
PhishTank & \multicolumn{1}{c|}{7.15\%} & 08:11 & \multicolumn{1}{c|}{22.38 \%} & 01:56 \\ 
OpenPhish & \multicolumn{1}{c|}{8.34\%} & 06:47 & \multicolumn{1}{c|}{37.19\%} & 01:32 \\ 
GSB & \multicolumn{1}{c|}{23.27\%} & 03:07 & \multicolumn{1}{c|}{84.13\%} & 00:49 \\ 
eCrimeX & \multicolumn{1}{c|}{12.40\%} & 05:24 & \multicolumn{1}{c|}{51.29\%} & 03:15 \\ \hline
\end{tabular}}
\end{table}

\begin{figure}[h!]
\centering
  \includegraphics[width=1\columnwidth]{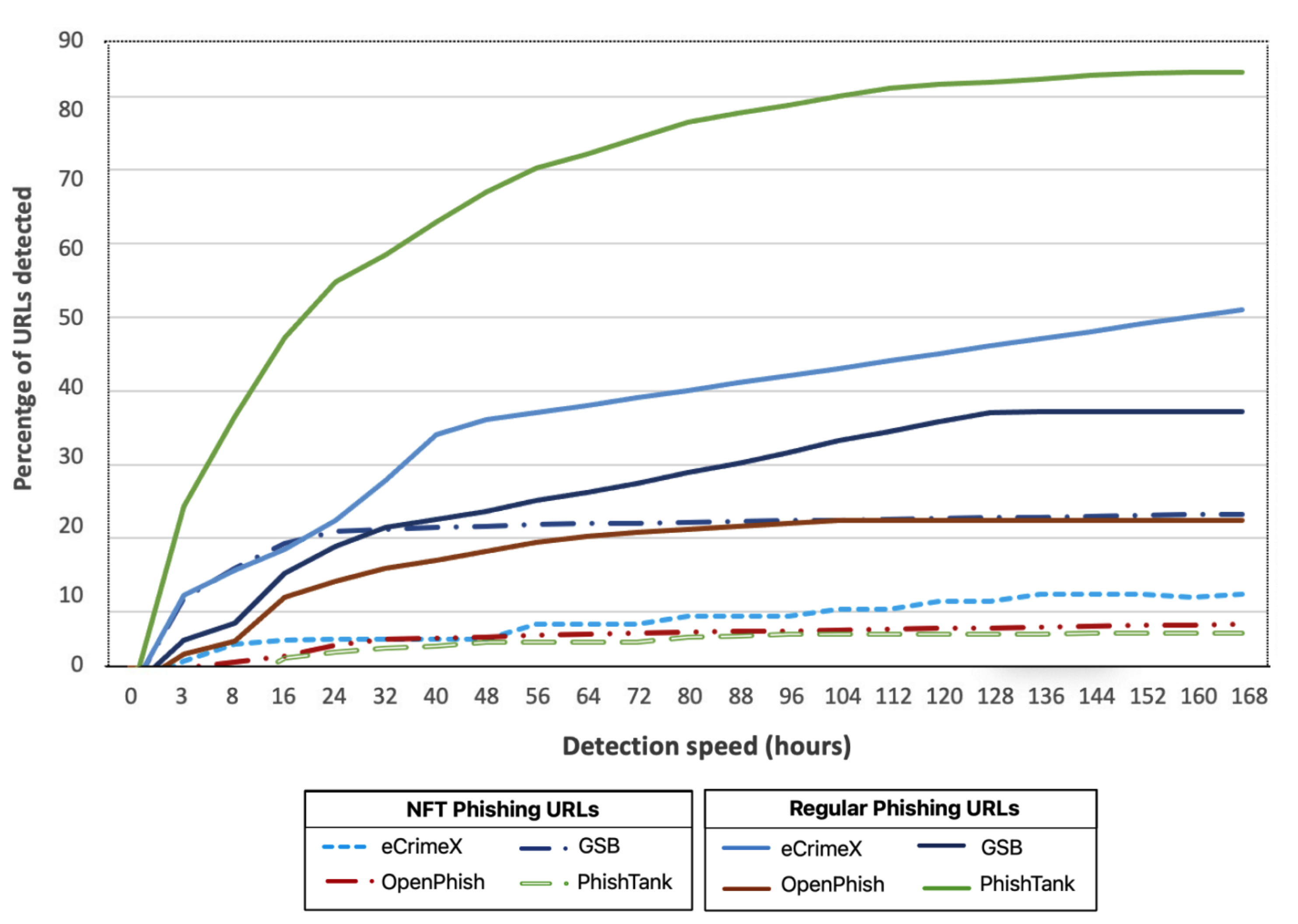}
\caption{Performance of anti-phishing blocklists against NFT-based phishing and regular phishing attacks. \sayak{Updated stats for new 382 URLs}}
  \label{fig:blocklist_coverage_overtime}
\end{figure}

\subsection{Browser Protection Tool Performance}  
\label{browser-tool}
\begin{figure}[h!]
\centering
  \includegraphics[width=0.9\columnwidth]{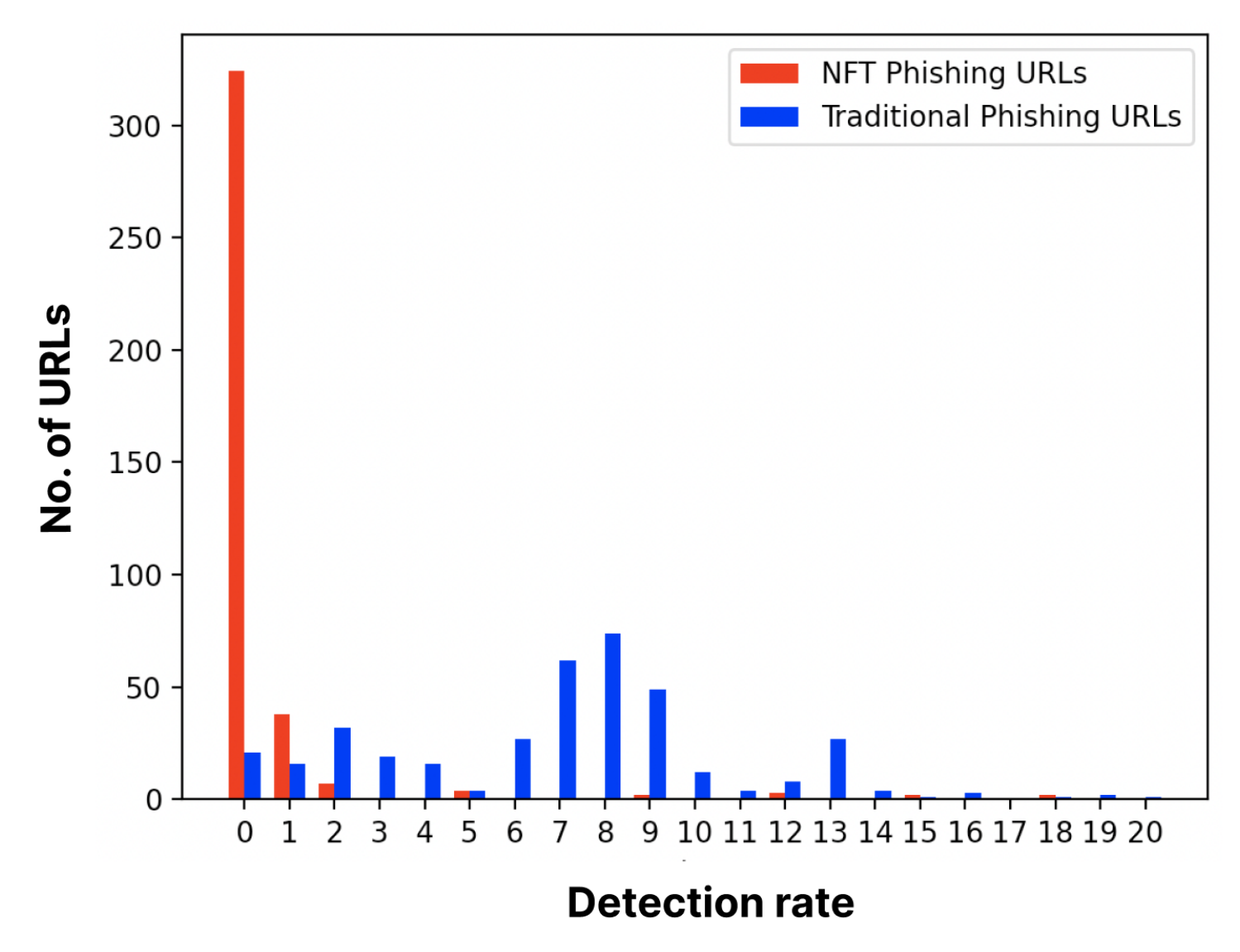}
\caption{Comparison of detection rates from browser protection tools for both NFT-based and traditional phishing attacks.}\sayak{Change: Now comparing NFT phishing with regular (traditional) phishing URLs}
  \label{fig:tools_coverage}
\end{figure}
Figure~\ref{fig:tools_coverage} shows the histogram of the detection rate of URLs found through our manual analysis (n=57) and through our detection tool (n=382). 
Nearly $74\%$ (n=$324$) of all NFT-based phishing URLs had \textit{zero} detections a week after their appearance in our dataset, with only $8.6\%$ (n=$38$) URLs having only \textit{one} detection. 
In comparison, regular phishing websites had a median detection rate of 6, a week after appearing in our dataset, with only $3.6\%$ of URLs (n=$16$) having \textit{zero} detection. When compared with the detection rate of the same number of randomly selected traditional phishing websites, we see only 19 (5\%) of them had \textit{zero} detections after a week, with detection rates distributed across various higher scores as shown in Figure~\ref{fig:tools_coverage}.
This further confirms that prevalent browser protection tools struggle significantly against NFT-based phishing attacks.  

For rug pull and pre-mint URLs, the findings were similar to the blocklisting performance in the previous section, with only four rug pull URLs having only \textit{one} detections throughout, while one URL had \textit{two} detections, whereas three of the pre-mint URLs had \textit{one} detection.



\subsection{Domain Detection}
\label{domain}
We found only $5$ NFT phishing websites that were removed by a domain provider over the course of a week, compared to $341$ URLs (~$78\%$) of regular phishing URLs.
Thus, to get a more substantial statistic for this analysis, we extended our analysis to three weeks.
Even after this longer period of time, only $61$ URLs (~$13.8\%$) of all NFT phishing URLs became inactive, and the median time of \textit{removal} for these URLs was $149$ hours ($6.2$ days). 
In comparison, $354$ regular phishing URLs (~$81\%$) became inactive, with a median removal time of $9$ hours. 
Moreover, we found 17 rug pull scams and 5 pre-mint scams that were removed throughout the duration of this analysis.
While it is possible that the domain had been removed due to malicious activity, it is more likely that the attacker did so after stealing the credentials/funds. 

\vspace{2mm}
\noindent
Thus, our findings highlight the significant gaps in existing anti-phishing measures for detecting and mitigating NFT-based scams. The disparities in detection rates and speeds between NFT phishing URLs and regular phishing URLs indicate that traditional anti-scam tools are not adequately equipped to sufficiently tackle the former at this stage. Thus, our automated detection tool can boost the current anti-phishing ecosystem toward detecting these attacks.

\section{Financial Impact of NFT Promotion Scams}
\sayak{Change: Merged financial impact for phishing+others. Included scams identified by our model}
\label{financial-impact-promoted-tweets}
\begin{table*}[]
\centering
\resizebox{0.9\textwidth}{!}{%
\begin{tabular}{|c|c|ccccc|ccccc|}
\hline
\multirow{2}{*}{Category} & \multirow{2}{*}{\# of} & \multicolumn{5}{c|}{Funds transferred (Approx)} & \multicolumn{5}{c|}{Transactions} \\ \cline{3-12} 
 &  & \multicolumn{1}{c|}{Total} & \multicolumn{1}{c|}{Min} & \multicolumn{1}{c|}{Max} & \multicolumn{1}{c|}{Mean} & Median & \multicolumn{1}{c|}{Total (approx)} & \multicolumn{1}{c|}{Min} & \multicolumn{1}{c|}{Max} & \multicolumn{1}{c|}{Mean} & Median \\ \hline
Phishing (Manual-TP) & 57 & \multicolumn{1}{c|}{\$804,294} & \multicolumn{1}{c|}{0} & \multicolumn{1}{c|}{\$176,701} & \multicolumn{1}{c|}{\$14,075} & \$2,194 & \multicolumn{1}{c|}{36,239} & \multicolumn{1}{c|}{2} & \multicolumn{1}{c|}{6100} & \multicolumn{1}{c|}{635.77} & 299 \\ 
Phishing (Automated-TP) & 382 & \multicolumn{1}{c|}{\$3,749,802
} & \multicolumn{1}{c|}{0} & \multicolumn{1}{c|}{\$401,754} & \multicolumn{1}{c|}{\$1,051} & \$745 & \multicolumn{1}{c|}{503,679} & \multicolumn{1}{c|}{0} & \multicolumn{1}{c|}{18,205} & \multicolumn{1}{c|}{1,318} & 328 \\ 
Rugpulls & 31 & \multicolumn{1}{c|}{\$385,391} & \multicolumn{1}{c|}{\$228} & \multicolumn{1}{c|}{\$210,213} & \multicolumn{1}{c|}{\$12,391} & \$10,894 & \multicolumn{1}{c|}{20,861} & \multicolumn{1}{c|}{4} & \multicolumn{1}{c|}{13,521} & \multicolumn{1}{c|}{672.93} & 259 \\ 
Pre-mint & 29 & \multicolumn{1}{c|}{\$47,221} & \multicolumn{1}{c|}{0} & \multicolumn{1}{c|}{\$45,698} & \multicolumn{1}{c|}{\$1,611} & \$15 & \multicolumn{1}{c|}{5,233} & \multicolumn{1}{c|}{2} & \multicolumn{1}{c|}{2,303} & \multicolumn{1}{c|}{180.44} & 192.5 \\ 
Legitimate & 145 & \multicolumn{1}{c|}{\$24,776,230} & \multicolumn{1}{c|}{0} & \multicolumn{1}{c|}{\$1,457,783} & \multicolumn{1}{c|}{\$170,264} & \$22,803 & \multicolumn{1}{c|}{639236} & \multicolumn{1}{c|}{3} & \multicolumn{1}{c|}{142182} & \multicolumn{1}{c|}{4408.5} & 1189 \\ \hline
\end{tabular}}
\caption{Descriptive statistics of funds transferred to the wallets belonging to fraudulent and legitimate NFT collections\sayak{Change: Added URLs caught by our model (Automated-TP)}}
\label{descriptive-stats-promotion-financial}
\end{table*}
To answer \textbf{RQ3}, we evaluate the financial impact of fraudulent NFT scams that were promoted on Twitter by looking at the number of cryptocurrency assets that were transferred to attacker-owned wallets. From our initial dataset of 823 NFT projects (where 300 projects were manually verified to be fraudulent), we were able to extract the wallet addresses for 37.2\% of all projects that are phishing (n=57), rug pulls (n=31) and pre-mint (n=29) and that belong to either the Ethereum or the Binance Smart Chain blockchain. We focus on these two blockchains due to their popularity and easily accessible public APIs.
We found over 62,333 transactions made to the wallet addresses attached to the fraudulent NFT projects \textit{during} or after the promotion occurred on Twitter. 
These transactions collectively transferred more than \$$1.24$ million to fraudulent wallets, with the median amount of funds sent to each of these wallets being about \$$2,590$. 
Nearly 73\% of all such transactions (n=3,819) indicated that the user had connected their wallet without any funds transferred.
As mentioned earlier, since some of these abandoned collections ask for full transaction rights to the user's wallet, using these credentials later on for malicious purposes is possible (and worth further exploration).

We also evaluated the financial impact of the phishing and pre-mint scams that were discovered by our automated detection model (our n=382 true positive detections). We found more than \$$3.74$ million that was transferred to attacker-owned wallets through these scams in 503,679 transactions. 
An overview of the transactions for each category of fraudulent NFT projects is provided in Table~\ref{descriptive-stats-promotion-financial}.
As a baseline, we also include the funds transferred to NFT projects that were identified as legitimate. Our findings indicate that fraudulent NFT projects that are promoted on Twitter can cause millions of dollars of damage with unsuspecting users transferring considerable amounts of money to attacker-owned wallets. Furthermore, the fact that users have connected their wallets without transferring funds, potentially granting full transaction rights to malicious actors, poses additional risks to the security of their digital assets. As seen earlier, Twitter has a slow response time and low coverage to remove posts that promote fraudulent NFT projects, and the virality of these posts over time can potentially expose many users to these scams. The lack of coverage of these scams by prevalent anti-scam tools and ML models further exacerbates the situation.
\section{Discussion and Conclusions}

\subsection{Negative Impact of NFT Promotions} Our work presents the first study on identifying the negative impact of artificially promoting  NFT collections on Twitter, with a large number 
of such projects being fraudulent.
Initial engagement by social bots drives engagement from real users, who are deceived into investing in these scams. 
Even our relatively modest dataset of identified scams showed more than \$5.1 million that was lost due to these attacks, indicating a much larger and more lucrative ecosystem for scammers. 
It is doubtful whether the accounts that promote these scams are unaware of their actions, as some of them are suspended not long after conducting promotions, with several others also deleting their old tweets to hide prior activity. These promotion-driven scams are another avenue for new attacks in an ecosystem that is already rife with scams. 
Given the lack of technical knowledge of many users with regards to the new, decentralized ecosystems, as well the poor coverage of these attacks by anti-scam tools (as evident from our measurement study Section~\ref{anti-phishing-effectiveness}), there is a need for social media sites to provide stricter moderation of such posts. Conversely, Twitter allows the promotion of content through sweepstakes to artificially increase user followers~\cite{twitter_contest_rules}, a strategy that can help the spread of scams on their platform.  We hope that our ML-based phishing classifier tool can help in counteracting these attacks on a large scale, and we look forward to contributing our work to the larger research community.

\subsection{Ethical Considerations}
For this study, we analyzed public tweets from Twitter (and phishing URLs from DNSTwist/Certstream for training our detection model in Section~\ref{ml-model-nft-phish}). 
For the former, we did not retain any identifiable information that can be attributed to the original poster, or information about users who had engaged with the tweet(s).
Similarly, we discarded all malicious URLs after our analysis period, and do not intend to distribute them. Any blockchain resources (such as contract addresses, and tokens) are already part of the public domain, due to the open nature of the ecosystem, but we again discarded any such identifiable resources including wallet transactions, minting history, etc.

\subsection{Limitations}
\subsubsection{Size of Dataset:}
In Section~\ref{collecting-nft-tweets}, we discuss how our preliminary manual analysis led us to use the ``NFT Promoter'' keyword to find the promotion accounts with at least 40k followers. We acknowledge that our dataset could have been expanded by considering additional promotional accounts that did not use this specific keyword. 
Also, we note that the (manual) task of identifying fraudulent NFT projects such as rug pull and pre-mint scams is generally very challenging, as they often bear a close resemblance to legitimate NFT projects, at least until the minting (or buying) stage, further requiring manual analysis to identify them.

However, despite the limited size of our dataset, we believe that it allowed us to 
get a good understanding of the modus operandi of NFT scams, their evasion tactics against anti-scam measures, and their financial impact. Our findings also informed the development of our detection classifier (Section~\ref{ml-model-nft-phish}), which was able to identify 382 new NFT phishing projects that were shared by 87 promotional accounts, 71 of which had not been previously identified in our initial analysis. 
Of these newly discovered accounts, only 18 did not include the ``NFT Promoter'' keyword in their profile description, suggesting that our classifier is not biased towards flagging accounts that contain this keyword only. We also did not find any new attack vectors related to the fraudulent NFT projects promoted by these accounts. We plan to utilize the findings of the classifier to recognize newer keywords, heuristics, and tweet templates, which can be used to detect newer fraudulent NFT projects in the future.
Given the fact that many fraudulent NFT projects and the accounts that promote them often remain undetected online for extended periods, our classifier can be a useful addition to improve the coverage and detection of these threats.

\subsubsection{Limited Blockchain Coverage:}Our analysis is primarily focused on projects that were hosted on the Ethereum and BSC (Binance Smart Chain) blockchains. These two platforms were selected due to their substantial contributions to the overall NFT market activity, thereby serving as representative samples for our study. Additionally, during our analysis period, both Ethereum and BSC offered public access to their transaction data and maintained well-documented APIs. These features were instrumental for our financial transaction examination and for extracting various on-chain transactional attributes necessary for our detection classifier model.
\shirin{should explain why we didn't look at rest.}\sayak{Added above} 
 As a follow-up to this work, we plan to expand our analysis towards NFTs hosted on other blockchains such as Solana~\cite{solana}, Polygon~\cite{polygon} etc.
\shirin{I would remove this paragraph unless the reviewers have asked for it.}
\sayak{Removed}

\bibliography{refs}
\end{document}